%Paper: astro-ph/9501110
%From: Franz Schoeniger <lschoen@c1.mtk.nao.ac.jp>
%Date: Tue, 31 Jan 95 20:42:04 JST

\def\tfr{{Tully-Fisher relation} }
\def\ref{\hangindent=1pc  \noindent}
\def\cen{\centerline}

\def\v{\vskip 2mm}

\def\no{\noindent}

\def\deg{$^\circ$}
\def\sect{\vskip 5mm \centerline}
\def\section{\vskip 5mm \centerline}

\def\deg{$^\circ$}

\def\title{\topskip 100pt \twelvepoint \bf \centerline}
\def\author{\topskip 50pt \vskip 3mm \tenpoint \rm \centerline }

\def\endpage{\vfil\break}
\def\cen{\centerline}

\cen{\bf The CO Tully-Fisher Relation for the Virgo Cluster }

\no
\cen{ Franz Sch\"oniger  \& Yoshiaki Sofue}

\no
\cen{\it Institute of Astronomy, University of Tokyo, Mitaka, Tokyo 181, Japan}

\no
{\bf Abstract.}

\v
\no
 A comparison between the CO and HI Tully-Fisher relation
for a sample of 30 Virgo ga\-la\-xies shows no significant difference
concerning the intrinsic scatter. The distance moduli from both
relations after correcting for the sample incompleteness
bias agree within the errors and also show no significant difference
with previous studies of the Virgo cluster using a complete
sample.
The CO total linewidth and the CO flux
were found to be  not well correlated. However,
if the distances to the individual galaxies
were calculated by the CO Tully-Fisher relation,
the CO luminosities calculated using these
distances are correlated
with  the CO linewidths. This correlation does
not show up  when the distances were calculated by the
conventional HI Tully-Fisher relation. Finally we give the
structure of the cluster as derived from this small subsample using
the distances to the individual galaxies. The large depth of the
cluster is confirmed both by the HI and the CO Tully-Fisher
relation.
\v
\no
{\bf Keywords:}  Galaxies: general\ -\ Distances: distance scale
  \ -\ Radio lines: 21 cm  \ -\ Radio lines: molecular

\v

\no
{\bf 1. Introduction}\v
\no
The rotational characteristics used in the \tfr
so far has mostly been determined
by observations of the 21 cm line of atomic hydrogen
(Tully \& Fisher 1977, Tully \& Fouqu{\'e} 1985, Aaronson et al 1986,
Pierce \& Tully 1988, Kraan-Korteweg et al. 1988; Fouqu{\'e} et al.
1990;
Fukugita et al. 1991). The use of CO instead of
HI was proposed by
Dickey \& Kazes
(1992). Due to the smaller beam in CO
compared to HI, the usage of CO could help to overcome the present limit
of 50 to 100 Mpc, up to which galaxies are reachable by observations in HI.
Dickey and Kazes found a good correlation between CO and HI linewidths
for the Coma  cluster and showed the hope that CO might
be a useful tool for distant galaxies. Sofue (1992) and Sch\"oniger \& Sofue
(1994) also found a good correlation between CO and HI for field galaxies,
despite most of the galaxies
in their sample showed peculiarities like interactions or active nuclei.
However, so far it has not been studied how the use of
CO affects the intrinsic scatter of the \tfr compared to
the use of HI. Besides this very important question concerning the reliability
and
the accuracy of the new method compared to the HI Tully-Fisher
relation any comparison between CO and HI linewidths as a distance indicator
enlarges our knowledge and may either support the applicability of the new
method
 or shows us its limitations.

\no
In this context and in order to study
cluster galaxies instead of field galaxies,
we compared the conventional blue \tfr for 30 Virgo cluster galaxies
with the corresponding CO Tully-Fisher relation.
We also looked for a possible correlation between
CO Flux and CO linewidth.  Since such a relation
might  be deteriorated by the depth effect of the cluster,
we corrected for the depth effect by calculating CO luminosities
from the   Tully-Fisher
distances.
Finally we compared the depth of the cluster along
the line of sight by using the Tully-Fisher distances
to the individual galaxies as derived from both
distance indicators.

\v
\no
{\bf 2. Data}\v

\no
All data used in our study was taken from the literature.
Most of the galaxies in our sample have been observed in CO
by Kenney\&Young (1988) and   Young
et al. (1985) using the 14 m antenna of the Five College
Radio Astronomy Observatory (HPBW 50''), some have been observed by
Stark et al. (1986) using the 7 m antenna at AT \& T Bell
Laboratories (HPBW 100'').
The limiting apparent blue magnitude corrected for inclination
and internal absorption of the  sample is
${\rm B_{T}^{0}=12.2}$. For the Tully-Fisher analysis we have used
the CO and HI linewidths at 20\% of peak level. Since the 50'' beam of
the FCRO antenna is too small to cover the CO emission of most of
the galaxies, we have integrated all the detected positions to a total line
profile. If the linewidth of the integrated profile was smaller than
the linewidth at 50\% of peak level as given in Stark et al.,
we have used the value from the BTL observations to avoid smaller CO
linewidths due to any under-sampling of the molecular gas.
The blue magnitudes used in the Tully-Fisher analysis were taken
from de Vaucouleurs et al. (1991). Inclinations
are from Huchtmeier \& Richter (1989) and HI linewidths are taken from
Bottinelli and Gougenheim (1991). A compilation of the galaxy properties
is shown in table 1.

\no
{\bf 3. Results}\v

\no
{\it 3.1. Tully-Fisher analysis for CO and HI}
\v\no
Fig. 1a shows a plot of CO versus HI linewidths for
the Virgo cluster galaxies. The correlation is obvious,
despite the scatter is larger than for our study of
field galaxies (Sch\"oniger and Sofue 1994).
In Fig. 1b we show a histogram of the linewidth ratio
${\rm W_{CO}/W_{HI}}$.
Here the tendency of the CO profiles to be narrower
than the corresponding HI profile can clearly be seen.
We get ${\rm W_{CO}/W_{HI}}=0.92\pm 0.13 $ compared to
 ${\rm W_{CO}/W_{HI}}=0.96\pm 0.1$ for our study of field galaxies.
  The question is whether the differences in linewidths between
CO and HI improve or deteriorate the Tully-Fisher relation
if CO  is used  instead of HI.

\no
For the  Tully-Fisher analysis the linewidths have to be corrected
for inclination {\it i}. Hereby usually low inclination galaxies  are rejected
in order to mimimize the error introduced by the uncertainty of the
inclination.
For a galaxy with
an inclination of 30\deg taken from a usually available catalogue
the uncertainty in {\it i} of $\pm 7$\deg
in inclination leads to an error of +0.7 mag and -0.52  mag respectively,
which can severely
influence a conclusion drawn from the Tully-Fisher relation.
In our study  we nevertheless don't reject low inclination galaxies.  The
reason will
be explained in detail later when we discuss the influence of the
sample incompleteness bias on our result.  For the moment we just point out
that we wanted to have a sample as complete as possible in linewidths.
For the sake of consistency with the
most recent and most complete study of the Virgo cluster in the
blue band (Fukugita et al.1993) we did not apply a correction
for internal turbulence.

\no
Fig. 2 shows the apparent blue magnitude versus HI linewidths corrected
for inclination
(Fig. 2a) and versus CO linewidths (Fig. 2b). Low inclination galaxies
(i<30\deg) are marked as open circles.
The slopes are quite gentle compared to the slope for the calibrators,
$-3.05$ for the CO and $-4.5$ for the HI Tully-Fisher relation.
The dispersion is 0.52 for the HI \tfr and 0.58 for the CO
Tully-Fisher relation. This means that due to  the
similar intrinsic scatter in case of CO and HI  it is difficult to
decide which relation gives  more accurate distances.
Excluding low inclination galaxies does not change this result.

\no
The HI Tully-Fisher relation was calibrated using six local
calibra\-tors with Ce\-pheid distances given by Ma\-dore
and Freed\-man (1991), yiel\-ding:
$${M_{B}}=-(6.31\pm 0.5) {\rm log} W_{i}-3.72\pm 1.24$$
Of the local calibrators only M 31 has been completely mapped in CO
and a total line profile is available. In case of M 31 the CO and HI profiles
agree perfectly. For the other local calibrators we apply the result from
our study for field galaxies (Sch\"oniger \& Sofue 1994) and adopt a CO total
linewidth being four percent smaller than the corresponding HI linewidth.
This yields a calibration for the CO Tully-Fisher relation of the
form:
$${M_{B}}=-(6.22\pm 0.5) {\rm log} W_{i}-4.07\pm 1.17 $$
Using this calibrations we get for the cluster
distance moduli  $(m-M)_{0}({\rm CO})=30.90\pm 0.70$ and
$(m-M)_{0}({\rm HI})=31.03\pm 0.52$. The smaller distance derived
from CO hereby reflects the tendency of the CO linewidths
to be narrower than the corresponding HI linewidths.

\v
\no
{\it 3.2. Correction for the sample incompleteness bias}
\v
\no
An important prerequisite to be able to derive a correct distance
from a Tully-Fisher analysis is its application to  a complete sample.
As pointed out by Teerikorpi (Teerikorpi 1984) any limiting  magnitude leads
to a more gentle slope in the Tully-Fisher relation and therefore yields
a distance smaller than the true value. This sample incompleteness bias or
Malmquist bias,
as it's also called sometimes, despite not fully correct, has plagued the
Tully-Fisher
studies for a long time and led for example  Aaronson to the conclusion that
the Tullly Fisher relation might be curved (Aaronson et al. 1986).

\no
In our case we have
a sample   complete down to a limiting magnitude of 12.2. This means
our result suffers from a severe sample incompleteness
bias. This results for example in the very gentle slope of the Tully-Fisher
relation compared to previous studies using a more complete sample which yield
a
much steeper  slope
between 6 and 8. It also leads to the unusually small values for the Virgo
distance modulus.  In order to correct for the effect of the sample
incompleteness
bias we follow a prescription by Schechter (Schechter 1980) who suggested that
an unbiased mean distance independent of the limiting magnitude
can be derived from the inverse  Tully-Fisher relation
if the  the coverage in log $\rm W_{i}$  is complete. Since we need a
completeness
in linewidths to apply this procedure we have not rejected low inclination
galaxies.
Fouque has demonstrated the applicability of this method
for the Virgo cluster (Fouque et al. 1990). Inverse Tully-Fisher relation
hereby
means that the linewidth is treated  as the dependent variable containing all
errors.
The slope derived this way then is forced onto the calibrators getting the
inverse Tully-Fisher relations as follws:

$${M_{B}}=-8.69\ {\rm log} W_{i}+2.09\pm 0.26  $$
in case of HI and
$$   {M_{B}}=-8.67\ {\rm log} W_{i}+1.92\pm 0.27                       $$
in case of CO.

\no
It can be seen that the slopes of CO and HI Tully-Fisher relation now are in
agreement.
Appliing these relations  we derive a distance modulus of $31.44\pm 0.70$   in
case of
HI and of   $31.18\pm 0.92$   in case of the CO Tully-Fisher relation.  The
result
from HI agrees  with the result  from Fukugita et al. (1993) who also derived
a distance modulus of 31.44 from a
complete sample. The result from CO still gives a somewhat smaller value,
whereas
the difference between the two distance moduli  even became larger than before
appliing the correction for the Malmquist bias. Nevertheless also the distance
modulus derived from CO increased by almost 0.3 mag due to the application
of the inverse relation.

\no
We also checked the effect of restricting our sample to galaxies with
inclinations
larger than 45\deg. From the remaining 20 galaxies both relations
yield
distance moduli 0.2 mag smaller than from the sample including all
galaxies.
This shows that in order to get a complete coverage in linewidths it is
actually
necessary to include all galaxis.

\no
Finally we  would like to stress again explicitly that despite our study is
conducted
on a sample which suffers from a severe incompleteness and possibly from
undersampling of the CO emission we were able to derive a distance modulus to
the
Virgo cluster from the CO observations which disagrees by only 10 percent  with
previous HI studies using by far larger samples. We can only speculate about
the
reason for the discrepacy itself, however we can say that it is of an order
which still supports an optimistic prospect concerning the future
applicability of the Tully-Fisher relation.

\no
This result is particularly important for future applications of the CO
Tully-Fisher relation to distant clusters of galaxies which can not be reached
by HI. Due to the difficulty  of the observations we can hardly expect to
be able to observe a complete sample, therefore a correction for
the Malmquist bias is essential. The use of the inverse Tully-Fisher relation
can help to overcome this problem.

\v
\no
{\it 3.3. CO Flux vs CO linewidth}
\v
\no
Kenney and Young (1988) in their paper also calculated the
total CO flux ${\rm S_{CO}}$ by fitting the measured intensities
at the detected positions to a brightness temperature distribution
convoluted with the telescope beam. CO flux and the observed
quantity $\int T_{A}^{\star}dv$ are related as follows:
$$
S_{CO}= {{2k}\over{\lambda ^2}} \int\int T_{Rav}(\Omega, v)\ d\Omega\ dv
$$
whereas $T_{Rav}(\Omega, v)$ is the average
brightness temperature within the main  beam and directly proportional
to the antenna temperature $ T_{A}^{\star}$.

\no
In Fig. 3 we show a plot of the CO flux vs the blue magnitude.
Despite the  scatter in this plot there is a
clear tendency of  galaxies being more luminous
in the blue having also  higher CO fluxes. This actually
should also  lead to
a correlation between CO flux  and CO linewidth.
In Fig. 4 we have  plotted
the CO flux versus the CO linewidth for our sample galaxies.
The correlation in this plot is not very good;
the correlation coefficient is $R=0.59$.
However, since any possible correlation of these quantities
could be disturbed by the depth effect of the Virgo cluster,
we corrected for the depth effect by plotting luminosities
instead of intensities.
We  calculated the luminousities first using the HI distances
and secondly using the distances derived from the CO Tully-Fisher
relation, and then plotted them separately versus the HI and  CO
linewidths.

\no
The result is shown in Fig. 5. The luminosities
derived from the CO Tully-Fisher distances are much better correlated
with the CO linewidth ($R=0.79$, Fig. 5a) than the
luminosities calculated from the
HI Tully-Fisher distances ($R=0.68$, Fig. 5b). The correlation of
the luminosities derived from HI with the HI linewidths ($R=0.67$,
Fig. 5c) is also worse than in Fig. 5a. The good correlation in Fig. 5a
is quite conspicious and fuels speculations about a possible
"self consistent CO Tully-Fisher relation", from which galaxy distances
can be derived using only CO observations. However, the correlation only shows
up when the CO Tully-Fisher distances are used and therefore first
has to be checked
also in case of  other clusters of galaxies.
Also it has to be stated that a correlation between CO luminosity
and CO linewidth could introduce a Malmquist bias onto the
inverse CO Tully-Fisher relation.

\v
\no {\it 3.4. The Virgo distance and depth}

\no
Finally we compare the distances to the individual galaxies
using the HI and the CO Tully-Fisher relation.
A compilation of the measured linewidths and the resulting
distances  can be seen in Table 2 and
Fig. 6a plots the resulting HI distances versus the corresponding CO distances.
The smaller CO distances hereby reflect the tendency of the CO linewidths to be
narrower than the corresponding HI linewidths.
The HI mean distance is 16.5 ($\sigma=4.2$) Mpc, while
CO value is 15.2 ($\sigma=5.7$) Mpc.
In Fig. 6b we show for comparison the  result for normal (non-cluster)
galaxies which is reproduced from
Sch\"oniger and Sofue (1994). Since the discrepancies are small compared
to the errors it is difficult to conclude  that the
behaviour of the cluster galaxies is fundamentally different from field
galaxies. However, the tendency to show smaller linewidths in CO than in HI
cannot be denied for our sample. For the Coma cluster Dickey and Kazes
(1992) found evidence for rather the opposite, therefore we have to discuss
the possible reasons for this discrepancy, as there were the influence of
the cluster environment, the possible undersampling of the CO
emission and a statistical fluctuation.

\no
The smaller CO linewidths compared to HI for our sample are quite
obvious.
However, there are only three galaxies which do not follow this
trend.
For these galaxies, NGC 4438, NGC 4579 and NGC 4651, CO gives
a  larger distance  than HI. These three
galaxies are extremely deficient
in HI (Kenney \& Young 1986)
and have almost completely lost their atomic gas due to ram-pressure
interaction with the intra cluster medium. The stripping of HI due
to ram pressure plays a significant role in the case of the Virgo cluster
and the three galaxies mentioned above are extreme examples of this
effect.
This means that the distances of galaxies which are severely
deficient in HI can easily be underestimated by HI and should
therefore better be determined
by CO. For the galaxies with the larger HI distances another
effect must be considered as the origin of the difference between
CO and HI linewidths, since ram pressure stripping rather leads to
smaller HI linwidths than to the larger HI linewidths we actually
found. Tidal interactions however can disturb the velocity field
of a galaxy in a manner which leads to a wider line profile.
Such kind of interaction might be the reason for larger HI linewidths
in some cases. However we  of course can neither exclude undersampling of
the CO emission as a possible reason nor that its just a statistical
effect, since within one sigma the linewiths and distances from HI and CO
agree.

\no
We also tried to derive
the cluster structure by calculating the distances to the
individual galaxies. For this purpose we of
course have to use again the direct Tully-Fisher relation which means
we are facing again the problem of a poor sample picking out only
very bright  spirals. For this analysis we exclude the low inclination
galaxies, since we don't need the complete coverage in linewidths.
Of course we cannot expect to derive a detailed structure of the Virgo
cluster from such a poor sample. We rather want to look wether the
density excess up to distances of about 30 Mpc also shows up in our sample
and therefore plotted a galaxy density-distance histogram similar
to Fukugita's study.

\no
Fig. 7a shows the glaxy density-distance histogram for the HI Tully-Fisher
relation
and Fig. 7b for the CO Tully-Fisher relation.
As it was found by Fukugita et al (1993),
from our small subsample we also can confirm that the individual distances
are largely scattered
due to the depth effect of the Virgo cluster both in CO and HI
determinations and that a density excess exists up to distances of around 30
Mpc. Any further statement about the differences between the two distributions
are unjustified due to the poor sample.

\v
\no
{\bf 4. Conclusion}\v

\no
The intrinsic scatter of CO and HI Tully-Fisher relation for 30
Virgo cluster galaxies turned out to be comparable.
The Tully-Fisher analysis yields distance moduli of
31.18$\pm 0.92$ for the CO \tfr and 31.44$\pm 0.70$ for the HI
Tully-Fisher relation after correcting for the effect of the sample
incompleteness
bias. Considering the poor sample and the ther possibly incomplete
coverage of the CO emission in some cases the result is in good
agreement with previous studies and fuels optimism
about  the possibility to apply the CO Tully-Fisher relation
to distant clusters of galaxies where only a small sample of galaxies
can be observed.
A  correlation between CO luminosity and CO linewidth
showed upe when appliing CO Tully-Fisher distances to derive
CO luminosities. The lack of the correlation in case of the application of the
HI
Tully-Fisher relation is conspicious and  therefore the correlation
has to be regarded with scepticism and has to be checked in the case of
other samples.
The depth effect of the Virgo cluster could be demonstrated using the
CO as well as the HI Tully-Fisher relation with a result
similar to what was derived from a complete sample.

\v
\no
{\it Acknowledgements.} F. Sch\"oniger  receives support
from the German Academic Exchange Service in the framework of
the Zweites Hochschulsonderprogramm HSPII/AUFE
\v\v

\sect{\bf References} \v

\ref Aaronson M., Bothun G., Mould J., Shommer
   R. A., Cornell, M. E., 1986, ApJ 302, 536

\ref Bottinelli L., Fouqu{\'e} P., Gouguenheim L.,
     Paturel G., 1990, A\&AS 82, 391 %HI catalog

\ref de Vaucouleurs G., de Vaucouleurs A., Corwin  H. G. Jr., et al.,
   1991, in
   {\it  Third Reference Catalogue of Bright Galaxies}
   (Springer Verlag, New York)

\ref Dickey J., Kazes I., 1992, ApJ 393, 530

\ref Fukugita M., Okamura S., Tarusawa K., et al., 1991, ApJ 376, 8

\ref Fukugita M., Okamurta S., Yasuda S., 1993, ApJ 412, L13

\ref Fouqu{\'e} P., Bottinelli L., Gouguenheim L., et al., 1990, ApJ 349, 1

\ref Huchtmeier W., K., Richter O.-G., 1989, in {\it A General Catalog
  of HI Observations of Galaxies}

\ref Kraan-Korteweg R. C., Cameron L. M., Tamann G. A., 1988, ApJ, 331,
     620

\ref Kenney J. D. P., Young J. S., 1986, ApJ, 301, L13

\ref Kenney J. D. P., Young J. S., 1988, AJS, 66, 261

\ref Madore B.F., Freedman W. L., 1991, PASP, 103, 933

\ref Pierce M. J., Tully R. B., 1988, ApJ, 330, 579

\ref Schechter  1980

\ref Sch\"oniger F., Sofue Y., 1994, A\&A 283, 21

\ref Stark A. A.,  Knapp G. R., Bally J.,  Wilson R. W.,
   Penzias A. A., Rowe H. E., 1986, ApJ 310, 660

\ref Sofue Y., 1992, PASJ 44, L231

\ref Teerikorpi  P., 1984, A\&A 141, 407

\ref Tully B., Fisher J. R., 1977, A\&A 64, 661

\ref Tully B., Fouqu{\'e} P., 1985, ApJS 58, 67

\ref Young J. S., Scoville N. Z., Brady E., 1985, ApJ 288, 487

\vfill
\eject
\no
{\bf Figure Captions}
\v
\no
{\bf Figure 1a and b.} CO vs HI linewidths ({\bf a}) and
             and a histogram of galaxy numbers vs $W_{CO}/W_{HI}$
             ({\bf b})
\v
\no
{\bf Figure 2a and b.} Blue apparent magnitudes plotted vs HI ({\bf a})
                          and CO ({\bf b}) linewidths, low inclination
galaxies (i<45\deg) are plotted as open circles
\v
\no
\no
{\bf Figure 3.} CO flux plotted against apparent blue magnitudes. Open circles
show low inclination galaxies (i<45\deg).
\v
\no
{\bf Figure 4.} CO flux plotted against CO linewidths. Open circles
show low inclination galaxies (i<45\deg).
\v
\no
{\bf Figure 5 a b and c.} CO luminosity ${\rm L_{CO}}$ calculated from the CO
Tully-Fisher
relation vs CO linewidth
({\bf a}), from the HI Tully-Fisher relation ({\bf b}) versus CO
linewidth, and from the HI Tully-Fisher relation vs HI
linewidth ({\bf c}). Open circles
show low inclination galaxies (i<45\deg).
\v
\no
{\bf Figure 6a and b.} CO Tully-Fisher distances plotted against
                         HI Tully-Fisher distances for our Virgo sample
                        ({\bf a}) and for our study of field galaxies
                          ({\bf b})
\v
\no
{\bf Figure 7a and b.} Histogram of galaxy density as a function of  the HI
Tully-Fisher distance ({\bf a}) and CO Tully-Fisher
distance ({\bf b})

\vfill
\eject
\nopagenumbers
$$\vbox {\rm \halign { #\hfil && \quad \hfil #\hfil \cr
\multispan4 Table 1: Galaxy properties for our sample galaxies \hfil & \cr
\noalign{\smallskip}\cr
\noalign{\hrule}\cr
\noalign{\smallskip}\cr
Galaxy & $i$ & ${\rm B_{0}^{T}}$ &

   ${\rm S_{CO}}$ \cr
       &  [$^\circ$]    &
       & [Jy km/s] \cr
\noalign{\smallskip}\cr
\noalign{\hrule}\cr
NGC 4064  & 67 & 11.84  & 93\cr
NGC 4192  & 74 & 10.07  & 940\cr
NGC 4212  & 47 & 11.37 & 510 \cr
NGC 4216  & 80 & 10.29  & 620  \cr
NGC 4237  & 46 & 12.20  & 624\cr
NGC 4254  & 28 & 10.08  & 3000\cr
NGC 4293  & 76 & 10.95  & 270\cr
NGC 4298  & 67 & 11.62 & 660\cr
NGC 4302  & 90 & 11.39 & 620\cr
NGC 4303  & 25 & 10.07  & 2280\cr
NGC 4312  & 78 & 11.83  & 160\cr
NGC 4321  & 28 & 9.95   & 3340\cr
NGC 4388  & 74 & 10.82  &  230\cr
NGC 4402  & 75 & 11.70  &  630\cr
NGC 4419  & 67 & 11.63  & 920\cr
NGC 4438  & 65 & 10.56  & 210\cr
NGC 4450  & 44 & 10.70  &  450\cr
NGC 4501  & 58 & 9.86   & 2220\cr
NGC 4527  & 72 & 10.65  &  1810\cr
NGC 4535  & 43 & 10.34  & 1570\cr
NGC 4536  & 67 & 10.56  & 740\cr
NGC 4548  & 36 & 10.80  &  540 \cr
NGC 4569  & 63 & 9.79   & 1500\cr
NGC 4571  & 27 & 11.75 & 380\cr
NGC 4579  & 36 & 10.23 & 910 \cr
NGC 4647  & 36 & 11.83 &   600\cr
NGC 4651  & 45 & 11.10 &  350\cr
NGC 4654  & 52 & 10.75 & 730  \cr
NGC 4689  & 30 & 11.37 &  710\cr
NGC 4698  & 55 & 11.26 &   90\cr
\noalign{\smallskip}\cr
\noalign{\hrule}\cr
}}$$

\vfill
\eject
\no

$$\vbox {\rm \halign { #\hfil && \quad \hfil #\hfil \cr
\multispan6 Table 2: Linewidths and resulting distances for our sample
 galaxies \hfil & \cr
\noalign{\smallskip}\cr
\noalign{\hrule}\cr
\noalign{\smallskip}\cr
Galaxy & $W_{20}$(CO) & $W_{20}$(HI) &
 ${\rm W_{CO}/W_{HI}}$ &   ${\rm D_{CO}}$ & ${\rm D_{HI}}$
   \cr
     &    [km/s] & [km/s] &    & [Mpc]
       & [Mpc]  \cr
\noalign{\smallskip}\cr
\noalign{\hrule}\cr
NGC 4064  & 165 & 199   & 0.83 &  8.9 & 11.3 \cr
NGC 4192  & 420 & 476   & 0.88 & 12.2 & 14.2 \cr
NGC 4212  & 263 & 274   & 0.96 & 17.3 & 18.2  \cr
NGC 4216  & 520 & 540   & 0.96 & 15.1 & 15.9    \cr
NGC 4237  & 240 & 263   & 0.91 & 23.1 & 25.9  \cr
NGC 4254  & 225 & 264   & 0.85 & 13.7 & 16.8  \cr
NGC 4293  & 240 & 381   & 0.59 &  8.2 & 15.9 \cr
NGC 4298  & 250 & 265   & 0.94 & 13.6 & 14.7  \cr
NGC 4302  & 315 & 377   & 0.84 & 14.8 & 18.5  \cr
NGC 4303  & 160 & 171   & 0.94 & 10.1 & 11.0 \cr
NGC 4312  & 190 & 219   & 0.87 &  9.8 & 11.8  \cr
NGC 4321  & 223 & 269   & 0.83 & 12.8 & 16.2  \cr
NGC 4388  & 350 & 381   & 0.92 & 13.6 & 15.2   \cr
NGC 4402  & 240 & 286   & 0.84 & 12.6 & 15.8   \cr
NGC 4419  & 360 & 286   & 1.26 & 21.7 & 16.2  \cr
NGC 4438  & 320 & 350   & 0.91 & 11.6 & 13.0  \cr
NGC 4450  & 180 & 306   & 0.59 &  8.4 & 16.4   \cr
NGC 4501  & 515 & 532   & 0.97 & 16.7 & 17.4  \cr
NGC 4527  & 395 & 388   & 1.01 & 14.9 & 14.6   \cr
NGC 4535  & 260 & 294   & 0.88 & 11.5 & 13.5  \cr
NGC 4536  & 310 & 336   & 0.92 & 11.0 & 12.1  \cr
NGC 4548  & 272 & 264   & 1.03 & 18.3 & 17.6    \cr
NGC 4569  & 350 & 360   & 0.97 &  9.3 & 9.7   \cr
NGC 4571  & 170 & 173   & 0.98 & 21.7 & 22.2  \cr
NGC 4579  & 400 & 363   & 1.10 & 23.8 & 21.1  \cr
NGC 4647  & 180 & 216   & 0.83 & 17.5 & 22.0   \cr
NGC 4651  & 420 & 390   & 1.07 & 28.8 & 26.2  \cr
NGC 4654  & 270 & 304   & 0.89 & 12.2 & 14.2  \cr
NGC 4689  & 200 & 197   & 1.02 & 19.8 & 19.4   \cr
NGC 4698  & 275 & 317   & 0.87 & 13.0 & 15.0  \cr
\noalign{\smallskip}\cr
\noalign{\hrule}\cr
}}$$

\endpage
\bye
\end